\documentclass[preprint,superscriptaddress,aps,longbilbiography]{revtex4-2}

\usepackage{graphicx}
\usepackage{dcolumn}
\usepackage{bm}

\usepackage[utf8]{inputenc}
\usepackage[T1]{fontenc}
\usepackage{mathptmx}
\usepackage{xcolor}
\usepackage{ulem}
\begin{document}


\title{Hybrid terahertz emitter for pulse shaping and chirality control}

\author{Weipeng Wu}
\affiliation{Department of Physics and Astronomy, University of Delaware, Newark, Delaware 19716, USA}%
\author{Wilder Acuna}
\affiliation{Department of Materials Science and Engineering, University of Delaware, Newark, Delaware 19716, USA}%
\author{Zhixiang Huang}
\affiliation{Department of Materials Science and Engineering, University of Delaware, Newark, Delaware 19716, USA}%
\author{Xi Wang}
\affiliation{Department of Materials Science and Engineering, University of Delaware, Newark, Delaware 19716, USA}%
\author{Lars Gundlach}
\affiliation{Department of Materials Science and Engineering, University of Delaware, Newark, Delaware 19716, USA}%
\author{Matthew F. Doty}
\email[]{doty@udel.edu}
\affiliation{Department of Materials Science and Engineering, University of Delaware, Newark, Delaware 19716, USA}%
\author{Joshua M. O. Zide}
\email[]{zide@udel.edu}
\affiliation{Department of Materials Science and Engineering, University of Delaware, Newark, Delaware 19716, USA}%
\author{M. Benjamin Jungfleisch}
\email[]{mbj@udel.edu}
\affiliation{Department of Materials Science and Engineering, University of Delaware, Newark, Delaware 19716, USA}%

\date{\today}

\begin{abstract}
\textbf{
Terahertz (THz) radiation, spanning from $0.3$ to $ 3\times10^ {12}$~Hz, fills the crucial gap between the microwave and infrared spectral range. THz technology has found applications in various fields, from imaging and sensing to telecommunication and biosensing. However, the full potential of these applications is often hindered by the need for precise control and manipulation of the frequency and polarization state, which typically requires external THz modulators. 
Here, we demonstrate a hybrid THz source that overcomes this limitation. Our device consists of two THz emitters integrated into one single device, enabling pulse shaping and chirality control of the emitted radiation without additional external components. The two sources are a spintronic emitter and a semiconductor photoconductive antenna (PCA). Using a combination of dual-wavelength excitation, allowing for control of the relative time delay between the two laser excitation pulses, and tuning external parameters for each emitter (i.e., biasing voltage for the PCA and magnetic field for the spintronic THz emitter) enables precise control of the mixing of the two signals and results in frequency, polarization, and chirality control of the overall THz radiation. 
This on-chip hybrid emitter provides an essential platform for engineered THz radiation with wide-ranging potential applications.
}


\end{abstract}

\maketitle

{Terahertz (THz) radiation, ranging from 0.3 THz to 30 THz in the electromagnetic spectrum, has been utilized across various applications such as imaging, sensing and spectroscopy {\cite{Jin2015}}, telecommunication {\cite{Ma2019}}, materials characterization, and biosensing {\cite{SpecAndImg2013, Cocker2013, dhillon20172017, FERGUSON20021043, Leitenstorfer_2023}}.
The non-destructive and non-ionizing properties of THz radiation are particularly interesting for industrial inspection and medical diagnosis applications {\cite{Pawar2013, Roh2022, Nakajima2007}}. Many of these applications benefit from precise control and manipulation of the frequency and polarization of the emitted THz radiation. For example, circular dichroism spectroscopy is an important tool for distinguishing between enantiomers of chiral drug molecules where the different chiralities can drastically alter the drug's interactions with bodily enzymes or proteins and lead to substantial differences in toxicity or pharmacological efficacy \cite{Mannschreck}. 
However, traditional THz sources rely on additional components such as external THz modulators that facilitate the manipulation of the THz radiation emitted from the source \cite{Wang_rev, Huang2024, Lee2017} to achieve control over properties such as circular polarization. Commonly, THz modulators manipulate the amplitude, frequency, or phase of incident THz waves. 
The polarization control is typically achieved by external means such as ion-gel gating \cite{Kim_Sci_Adv_2017} or the creation of photonic meta-materials-based microelectromechanical systems {\cite{Kan_APL_2013, Kan_Nat_Com_2015, Konishi2020, Markovich2013, Nathaniel2013, Wang2021,Huang2024}}. Most THz polarization modulation methods typically limit the bandwidth and frequency of the THz signal output. Therefore, efficient THz sources with direct and tunable control over the THz spectrum and polarization are urgently needed {\cite{JAP_wu_stripe, Khusyainov2021, APL_Wu_disks, Kong2019, Nathan2017, Kolejak2022, Khusyainov2021}}.

Here, we demonstrate a hybrid THz source that combines the distinct characteristics of two THz emitters integrated into one single device, enabling pulse shaping and chirality control of
the emitted radiation without the need for additional external components. For this purpose, we take advantage of the distinct material properties of a spintronic emitter (SE) and a traditional photoconductive antenna (PCA) and exploit their unique responses to femtosecond laser excitation at different wavelengths: the pump-wavelength sensitivity of the semiconducting material in the PCA and the pump-wavelength insensitivity of the spintronic emitter. Furthermore, the hybrid emitter benefits from the combined broadband emission characteristics of the spintronic THz generation process \cite{Seifert_2016} and the high-intensity emission property of the photoconductive switch below 1 THz \cite{Kong2019}. The output THz wave is created by combining the two individual THz pulses, enabling unprecedented pulse shaping and chirality control without any external modulators. The emission of the hybrid emitter can be easily controlled by addressing each of the two emitters (spintronic and PCA) independently using a biasing magnetic field (spintronic) and biasing voltage (PCA). Additional tunability of the polarization and manipulation of the frequency response is achieved by the relative delay between the two optical excitation pulses using a translation stage (see Fig.~\ref{fig1}). We demonstrate that the proposed emitter can be easily switched between linearly and circularly polarized THz radiation. Furthermore, the THz spectrum can be manipulated in real time. Therefore, our hybrid THz emitter offers intriguing opportunities for THz pulse engineering with wide-ranging potential technologies and applications.

The hybrid emitter comprises a PCA and a spintronic THz emitter. 
For traditional semiconductor-based PCAs, the emitted THz wave is determined by the material properties of the active semiconducting material, on which the photoconductive antenna is patterned for applying or detecting a voltage across two electrodes. The typical separation between the electrodes lies in the micrometer range.
The generation of charge carriers by femtosecond optical pulse excitation and their {subsequent rapid nonradiative decay} 
results in an ultrafast transient electrical current on picosecond time scales across the gap between biased electrodes. This transient is known as a Hertzian dipole. In the far-field approximation, the radiation from the dipole is given by {\cite{Jackson1998}}:
\begin{equation}
{E}_\mathrm{THz} \propto \frac{dI_\mathrm{PC}(t)}{dt},
\label{eq: PCA rad}
\end{equation} }
where $E_\mathrm{THz}$ is the far-field THz electric field and $I_\mathrm{PC}$ is the photocurrent generated in the gap. We note that (1) because {$E_\mathrm{THz}\propto V_\mathrm{bias}$, the THz power scales with the square of the applied biasing voltage  $V_\mathrm{bias}$}, (2) the generation of a THz electric field depends on the wavelength of the exciting optical pulse because charge carriers will only be created if the exciting pulse has photon energy larger than the semiconductor bandgap, and (3) the polarization of the emitted THz pulse is determined by the orientation of the bias voltage that determines the direction of propagation for the optically-generated electrons.

The THz generation process of the spintronic THz emitter is fundamentally different as it relies on the additional spin degree of freedom of electronic charge carriers \cite{Kampfrath2013,Wu2021}. This process is commonly understood as a laser-driven diffusive spin-polarized electron current generated in a ferromagnetic layer on picosecond time scales that is injected into an adjacent heavy-metal layer where the spin current is converted into a charge current by the inverse spin Hall effect:
\begin{equation}
\vec{J} _c \propto \gamma \vec{J}_s \times \vec{\sigma},
\label{Eq:ISHE}
\end{equation}
{where $\vec{\sigma}$ is the spin-polarization vector, $\gamma$ is the spin-Hall angle of the heavy metal layer, and $\vec{J}_\mathrm{c}$, $\vec{J}_\mathrm{s}$ are the charge and spin currents, respectively. In the far-field approximation, this oscillatory charge current results in THz radiation \cite{Wu2021}. The generation of the photo-excited diffusive spin current is independent of the wavelength of the optical pulse \cite{Papaioannou_2018}. The polarization of a THz wave emitted from a spintronic emitter can be controlled by the spin-polarization vector $\vec{\sigma}$, i.e., the macroscopic magnetization of the single-domain ferromagnet in which the spin current is generated.}

\section*{Results}
{The spintronic-PCA hybrid emitter is fabricated on a GaAs substrate using the following process: (1) growth of a GaAs film including embedded self-assembled ErAs nanoparticles which creates semiconducting material of thickness 2 $\mu$m with a bandgap of 1.55 eV (800 nm), (2) patterning of an \textit{I}-shaped photoconductive antenna [dimensions: $g=5~\mu$m, $s= 2.64$~mm, $w=10~\mu$m, see Fig.~\ref{fig2}(a)], (3) Growth of a spintronic emitter comprising a trilayer Ta(3~nm)/CoFeB(2~nm)/Pt(3~nm) on the opposite side of the substrate (see Methods for more information about materials growth and device fabrication). The sample is mounted on a PC board and attached to a hyper-hemispherical silicon lens to {better collect} the THz radiation {by overlapping the focal point of the off-axis parabolic mirror and the silicon lens}. A schematic of the hybrid emitter is shown in Fig.~\ref{fig2}(a). More details on sample fabrication are given in the Methods section.}

{All THz emission experiments are conducted using a standard time-domain THz spectroscopy (TDTS) system with a commercial PCA as the THz detector; see Fig.~\ref{fig1}. The laser is a frequency-doubled ultrafast Erbium fiber laser system with two mechanically switchable outputs at 780 nm or 1560 nm that are combined using a dichroic mirror (see Methods). The excitation beam is incident on the PCA face of the hybrid emitter; see inset in Fig.~\ref{fig1}. While the PCA is excited by the laser excitation pulse with 780 nm wavelength, it is not excited at  the laser excitation at 1560 nm because this wavelength corresponds to an energy below the bandgap {of the GaAs substrate} (1.42 eV – 870 nm). This means that the 780 nm excitation pulse is completely absorbed in the PCA and only the 1560 nm excitation pulse reaches the magnetic heterostructure deposited on the opposite side of the substrate. The spintronic emitter is pumped by the 1560 nm beam that passes through the semiconducting layer and substrate.}

Because the two sources are sufficiently close (500~$\mu$m) and the emitted THz waves propagate collinearly, they can be approximated as a single point source that generates a single THz pulse output whose properties depend on the mixing of the THz transients. 
We control this mixing by controlling the polarization of the emission from each of the two sources. The THz emission from the spintronic source is always polarized in the $\pm x0$-direction in the reference frame defined in Fig.~\ref{fig1}, which originates from application of an in-plane magnetic field ($\approx$1000 Oe) applied along the $\pm y0$-axis [see Eq.~(\ref{Eq:ISHE})]. The THz emission from the PCA source is polarized either parallel / antiparallel to that of the spintronic source (i.e.,~along the $\pm x0$-direction) or perpendicular to that of the spintronic source (i.e.,~along the $\pm y0$-direction), which originates from the bias voltage in the PCA applied along the $\pm x0$ (parallel / antiparallel) or $\pm y0$ (perpendicular) direction. When measuring the mixed wave from the hybrid source in the parallel or antiparallel configuration, we use a commercial PCA detector with electrodes oriented along the $x0$-direction, which is parallel (antiparallel) to the polarization of both sources. When measuring the mixed wave in the perpendicular configuration we use a pair of wire-grid THz polarizers (WGPs) in addition to the detecting PCA, as shown in Fig.~\ref{fig1}. WGP2, which is closest to the detecting PCA (see Fig.~\ref{fig1}), transmits THz waves with a polarization parallel to the $x0$-direction, the orientation of the gap electrodes in the detector. WGP1 (located closer to the hybrid emitter) is rotated by an angle of $\pm45^\circ$ with respect to WGP2. This allows for decomposition of the mixed THz pulse along two orthogonal directions, which define the $x1$-direction and $y1$-directions; see Fig.~\ref{fig4}(a). The THz transient electric field traces for different magnetic field and bias voltages are measured with a time resolution of 10 fs. We can easily isolate the contributions from the two sources by simply blocking one of the excitation pulse pathways (e.g.,~780~nm) to turn off the emission from the source (e.g.,~PCA) selectively driven by that excitation pulse wavelength.  {See Figs.~\ref{fig2}(b,c) and Fig.~S1 of the Supplementary Information (SI).}

\section*{Discussion}

 {Figures~\ref{fig2}(b,c) show the individual time-domain traces from the PCA and the SE when the 1560~nm beam is blocked [only the PCA emits THz radiation, Fig.~\ref{fig2}(b)] and the 780~nm beam is blocked [only the SE emits THz radiation, Fig.~\ref{fig2}(c)].} 
{The polarity of the THz signal from the PCA can be switched by reversing the biasing voltage from $V_\mathrm{bias}(+)$ to $V_\mathrm{bias}(+)$; see Fig.~\ref{fig2}(b). On the other hand, the polarity of the THz signal emitted by the SE can be switched by reversing the magnetization of the CoFeB layer using the external magnetic field [from $H(+)$ to $H(-)$]; Fig.~\ref{fig2}(c). This bias voltage and magnetic field dependence clearly demonstrate the different origins of each THz pulse emitted from the two emitters integrated into the hybrid source. Additional confirmation of the origin of the observed THz pulses is obtained by measuring the magnetic field dependence and bias voltage dependence of the THz radiation; see SI.}

 {Figure~\ref{fig2}(d) shows the key result for the parallel configuration: it provides evidence of pulse shaping and control over the relative orientation. Wire-grid polarizers are not used in the measurement because both of the THz signals contribute to the mixed signal are linearly polarized in the $x0$-direction}. 
{In the lower two traces (red/blue solid lines) the externally applied magnetic field stays the same and the THz emission from the SE always has an $H(+)$ character. The red and blue traces show that the emission from the PCA can be switched between $V(+)$ (red) and $V(-)$ (blue) by reversing the biasing voltage ($V_\mathrm{bias}$). In the upper two traces, the PCA source is kept in the $V(+)$ configuration and the SE signal is switched between $H(+)$ and $H(-)$ by reversing the magnetization ($M$) of the ferromagnetic material using the externally applied biasing magnetic field.}  {This results clearly shows that the pulse shape and the  orientation 
can be controlled using the applied magnetic field ($H$) and the biasing voltage ($V_\mathrm{bias}$) to separately address the SE and the PCA.}

{Figure~\ref{fig3}(b) (parallel configuration) demonstrates the pulse shaping that can be implemented with control over the relative time delay $\Delta T$ between the pulses exciting the PCA and SE (Fig.~\ref{fig1}). In the lowest (black) trace, for $\Delta T = -2$~ps, the THz pulses from the two sources are relative well separated in time and are thus essentially independent. As the delay time is varied, the THz pulse shape changes considerably, as shown by the other traces in Fig.~\ref{fig3}(b). The Fourier-transformed THz spectrum corresponding to each trace is shown in Fig.~\ref{fig3}(c). As is obvious from the figure, the superposition of the two pulses leads to a modification of the spectrum. The results shown in Figs.~\ref{fig3}(b,c) thus confirm that the controlled mixing of the two THz signals can be used to tailor the pulse shape in the time
domain and the THz spectrum in the frequency domain. A summary of the time-domain pulses and the corresponding FFT spectra for all possible combinations of the magnetic field and bias voltage directions is presented in videos in the SI.}

{Polarization and chirality control can be achieved in the perpendicular configuration as shown in Fig.~\ref{fig4}.  {The THz traces were measured in the $y1$-direction with opposite bias voltage polarities ($\pm 4$ V) in the $y0$-direction or the opposite magnetic field directions ($\pm 1000$ Oe) in the $x0$-direction.}
Conceptually, the polarization and chirality control originate from the fact that (1) the relative orientation of the THz electric field $\vec{E}$ of the PCA and SE THz emission can be controlled by changing either the applied magnetic field direction ($H$) or the biasing voltage ($V_\mathrm{bias}$) and (2) the phase delay between the two contributions to the mixed signal can be controlled by the pump time delay $\Delta T$ (Fig.~\ref{fig1}). To experimentally verify control over the mixed signal polarization and chirality, we use two wire-grid polarizers to measure the orthogonal components of the total electric field [see Fig.~\ref{fig1} and Fig.~\ref{fig4}(a)]. Figures~\ref{fig4}(b) and (c) show two representative THz waveforms with different chiralities: (b) left-handed and (c) right-handed. These traces are taken for the same magnetic field and bias voltage conditions and use a change in the time delay $\Delta T$ to control the phase delay between the PCA and SE contributions. Control over the chirality via changes to the applied field direction and bias voltage at a fixed time delay is shown in videos in the SI.


{To further support our experimental observations, we carried out numerical simulations of the mixed signal that results from the mixing of the two THz waves originating in the PCA and SE contributions (see SI). We started with the two individually measured THz pulses from the PCA and spintronic emitter. We then numerically simulated the resulting mixed signal by numerically superposing these two contributions in both the parallel and perpendicular configurations with variable time delay between the pulses in the simulations. As is shown in the SI, the simulations reproduce the experimental results remarkably well.  }

We note that the concept of pulse shaping and chirality control was demonstrated here by using a hybrid spintronic/semiconducting THz emitter. However, the idea extends to any system with two or more THz radiation sources that fulfill the following two conditions: (1) The THz sources are sufficiently close to one another (ideally fabricated on the same chip). (2) The two or more THz sources integrated into the hybrid emitter excite THz waves when pumped with different wavelengths, {which allows control over the phase delay between each contribution to the mixed THz output of the device}.  {We envision this could also be achieved by integrating two PCAs in a single emitter with different band engineered semiconductors, which can be excited at different wavelengths \cite{Acuna2024}}. We further note that straightforward modifications to the electrode design for the semiconductor PCA THz source will allow for switching between all parallel and perpendicular modes of operation without physical modifications to the experimental setup.

We proposed and realized a hybrid THz source {that provides control over both pulse shape and polarization} 
{by taking advantage of the distinct characteristics of two THz emitters integrated into a single device. This hybrid emitter takes advantage of the unique material properties of spintronic and semiconducting components in two ways. First, the pump-wavelength sensitivity of the semiconducting material in the PCA and the pump-wavelength insensitivity of the ferromagnetic heterostructure enable independent control over the relative phase of the THz emission from each component via a dual-wavelength excitation scheme. Second, the different response of the constituent materials to external stimuli (applied bias, magnetic field) provides control over the relative orientation of the THz pulse emitted from each component. We demonstrated that a device designed to exploit these material properties allows for convenient control over polarization (e.g.,~switching between linear and circular THz emission). Furthermore, we showed that the THz spectrum can be manipulated without changing any optical components (e.g.,~switching between two pulse chiralities). The control over the mixed signal that results from combining the two contributions of the two individual and independently-controlled THz pulses generated in two distinct material constituents enables unprecedented pulse shaping and chirality control without any external modulators.}

\section*{Online Methods}
\subsubsection*{Time-domain THz spectroscopy measurements}
The experimental THz emission measurements are conducted using a standard time-domain THz spectroscopy (TDTS) system with a commercial photoconductive antenna as the THz detector (BATOP, model number: bPCA-100-05-10-800-h-I); see Fig.~\ref{fig1}. The laser is a frequency-doubled ultrafast Erbium fiber laser system by TOPTICA Photonics, Inc. The laser outputs time-synchronized pulses at both 780 nm and 1560 nm. For the primary 780 nm ($\pm$ 10 nm) output, the average output power is > 80 mW with a pulse width of < 100 fs at 80 MHz. The secondary 1560 nm output is measured to have a pulse width of 72 fs with an average output power of >160 mW. To co-excite the hybrid emitter, the 780 nm beam is combined with the 1560 nm beam using a dichroic mirror. While the semiconducting material at the gap of the PCA is excited by the 780 nm light, it is transparent to the laser beam of 1560 nm. The spintronic emitter is then pumped by the 1560 nm beam that passes through the semiconducting layer and substrate. We can isolate the contributions from the two sources by  blocking one of the excitation pulse pathways (e.g.,~780~nm) to turn off the emission from the source (e.g.,~PCA) selectively driven by that excitation pulse wavelength. In the parallel configuration, we use the commercial PCA detector with electrodes oriented along the $x0$-direction, which is parallel (antiparallel) to the polarization of both sources. When measuring the mixed wave in the perpendicular configuration we use a pair of wire-grid THz polarizers (WGPs) as shown in Fig.~\ref{fig1}. WGP2, which is closest to the detecting PCA, transmits THz waves with a polarization parallel to the $x0$-direction, the orientation of the gap electrodes in the detector. WGP1 (located closer to the hybrid emitter) is rotated by an angle of $\pm45^\circ$ with respect to WGP2. This allows for decomposition of the mixed THz pulse along two orthogonal directions, which define the $x1$-direction and $y1$-directions; see Fig.~\ref{fig4}(a). The THz transient electric field traces are measured with a time resolution of 10 fs. All measurements are performed in a purged N$_2$ environment.

\subsubsection*{Sample fabrication}
The spintronic-PCA hybrid emitter is fabricated using a two-step process on GaAs substrate: (1) growth of a GaAs film with self-assembled ErAs nanoparticles as semiconducting material (thickness: 2 $\mu$m) with a bandgap of 1.55 eV (800 nm), followed by patterning of an \textit{I}-shaped photoconductive antenna. (2) Growth of a spintronic emitter comprising a trilayer Ta(3~nm)/CoFeB(2~nm)/Pt(3~nm) on the opposite side of the substrate. The sample is mounted on a PC board and attached to a hyper-hemispherical lens to collimate the THz radiation. A schematic of the hybrid emitter is shown in Fig.~\ref{fig2}(a).

The details of the growth and fabrication are presented in the following:

\textit{(1) GaAs film with self-assembled ErAs nanoparticles:} 
ErAs:GaAs material is grown by molecular beam epitaxy at an ultra-high vacuum (10$^{-10}$ Torr) on a semi-insulating (001) GaAs substrate. When Er is incorporated above the solubility limit, it self-assembles as nanoparticles of a few nm diameter. We use a growth rate of 1$\mu$m/h with an arsenic overpressure for GaAs and co-deposit Er to obtain approximately 0.5\% ErAs composition at 530$^\circ$C real substrate temperature measured by band edge thermometry. These ErAs nanoparticles with a rock salt structure act as recombination centers, decreasing the carrier lifetime to sub-picosecond values and pins the eﬀective Fermi level in the host material bandgap \cite{Acuna2024}. This is required for materials used in photoconductive switches targeting THz pulse emission and detection. Reducing the carrier lifetime with ErAs has the advantage of having a lower impact on carrier mobility than alternatives such as low-temperature GaAs (LT-GaAs) \cite{Ohara_2006}. Furthermore, this low ErAs composition does not impact the material bandgap (1.42 eV – 870 nm), making it compatible with 780 nm excitation and transparent to 1550 nm, which is required for the present work. After the growth of the semiconducting material, an \textit{I}-shaped photoconductive antenna is patterned using optical lithography using a positive photoresist and laser writing, followed by deposition of 10 nm Ti and 100 nm  Pt by electron-beam evaporation and then lift-off. The dimensions are $g=5~\mu$m, $s= 2.64$~mm, $w=10~\mu$m, see Fig.~\ref{fig2}(a). The large separation ($s$) between the electrodes reduces contributions to the THz radiation from outside the gap area ($g$). The relatively small gap of the active area ensures a wide spectral width, enabling an efficient mixing of the THz radiation from both sources so that the overall signal is broadband.

\textit{(2) Spintronic heterostructure:} 
To increase the intensity of the emitted THz radiation from the spintronic component of the hybrid emitter, a trilayer of Ta, CoFeB, and Pt is deposited on the opposite side of the substrate with the following sequence and thickness: substrate//Ta(3~nm)/CoFeB (2~nm)/Pt(3~nm); see Fig.~\ref{fig2}(a). Ta and Pt have opposite spin Hall angles, enhancing the THz signal \cite{Seifert_2016}; see SI. The three layers are grown using DC magnetron sputtering in an Ar atmosphere at 4.5 mTorr and 20 sccm gas flow (base pressure $< 10^{-7}$ Torr) without breaking the vacuum. 

\bibliography{aipsamp}

\section*{Data Availability}

{The data that support the findings of this study are available from the corresponding
author upon reasonable request.}

\section*{Code Availability}

{Computer codes for the simulations are available from DOI.}

\begin{acknowledgments}

{The authors thank Subhash Bhatt and Prof. John Q. Xiao, Department of Physics and Astronomy, University of Delaware, for assisting in the growth of ferromagnetic heterostructures and valuable discussions.}
This research was primarily supported by NSF through the University of Delaware Materials Research Science and Engineering Center, DMR-2011824. 
The authors acknowledge the use of the Materials Growth Facility (MGF) at the University of Delaware, which is partially supported by the National Science Foundation Major Research Instrumentation under Grant No. 1828141 and UD-CHARM, a National Science Foundation MRSEC, under Award No. DMR-2011824.

\end{acknowledgments}

\section*{Author contribution}

{M.F.D, J.M.O.Z., and M.B.J. conceived the idea of this study. W.W. and W.A. fabricated the samples. W.W. conducted the TDTS measurements. Z.H. and W. W. performed the simulations. All authors analyzed the data, discussed the results, and contributed to the writing of the manuscript.}

\section*{Competing Interests}

{The authors declare no competing interests.}

\newpage
\section*{Figures}
\begin{figure}[h]
    \centering
    \includegraphics[width=1\columnwidth]{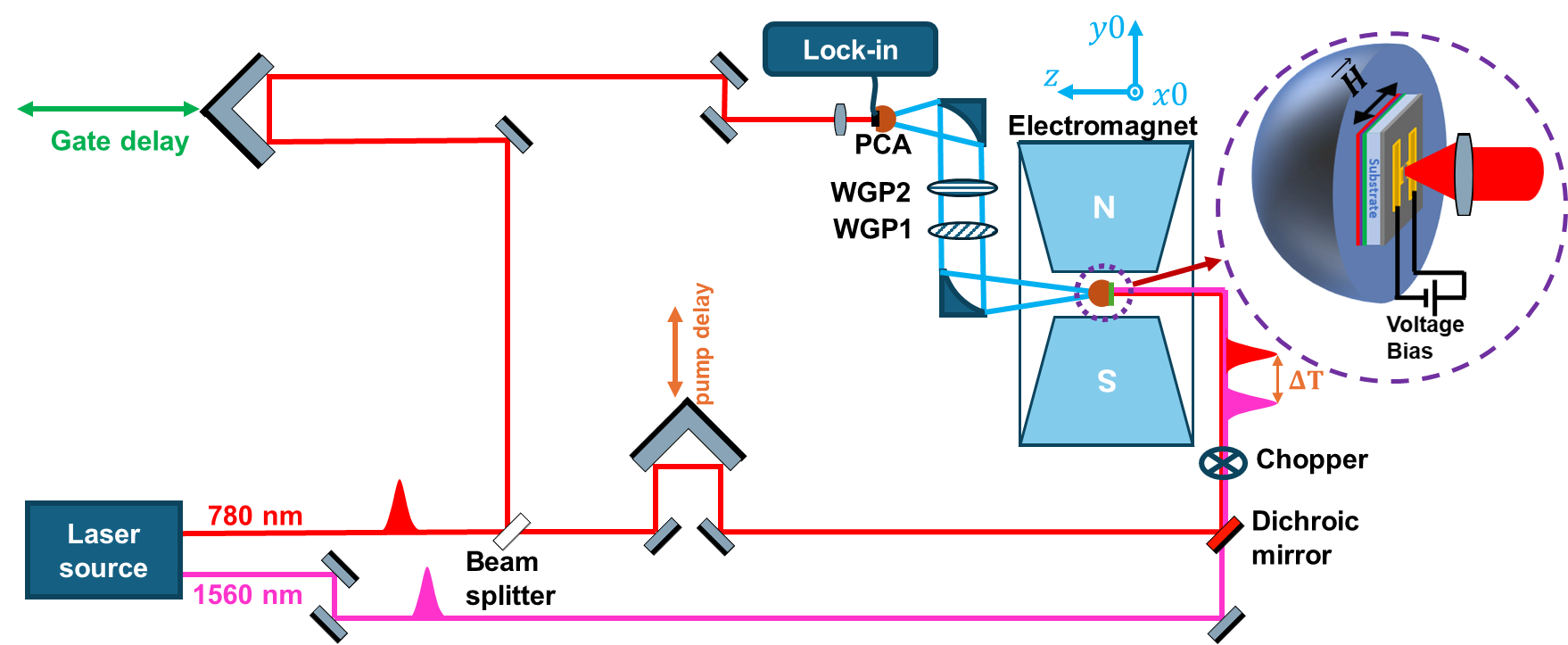}
    \caption{Schematic illustration of a time-domain THz spectroscopy (TDTS) system with a commercial photoconductive antenna (PCA) as THz detector. The time-domain traces of the THz signal are recorded using the gate delay stage, while the relative time difference between the 1560 nm and 780 nm pump beams is controlled by the pump delay stage. Two wire-grid polarizers (WGP1 and WGP2) are used to deconvolve the two orthogonal components of the THz signal.}
    \label{fig1}
\end{figure}
\newpage

\begin{figure}[t]
    \centering
    \includegraphics[width=0.75\columnwidth]{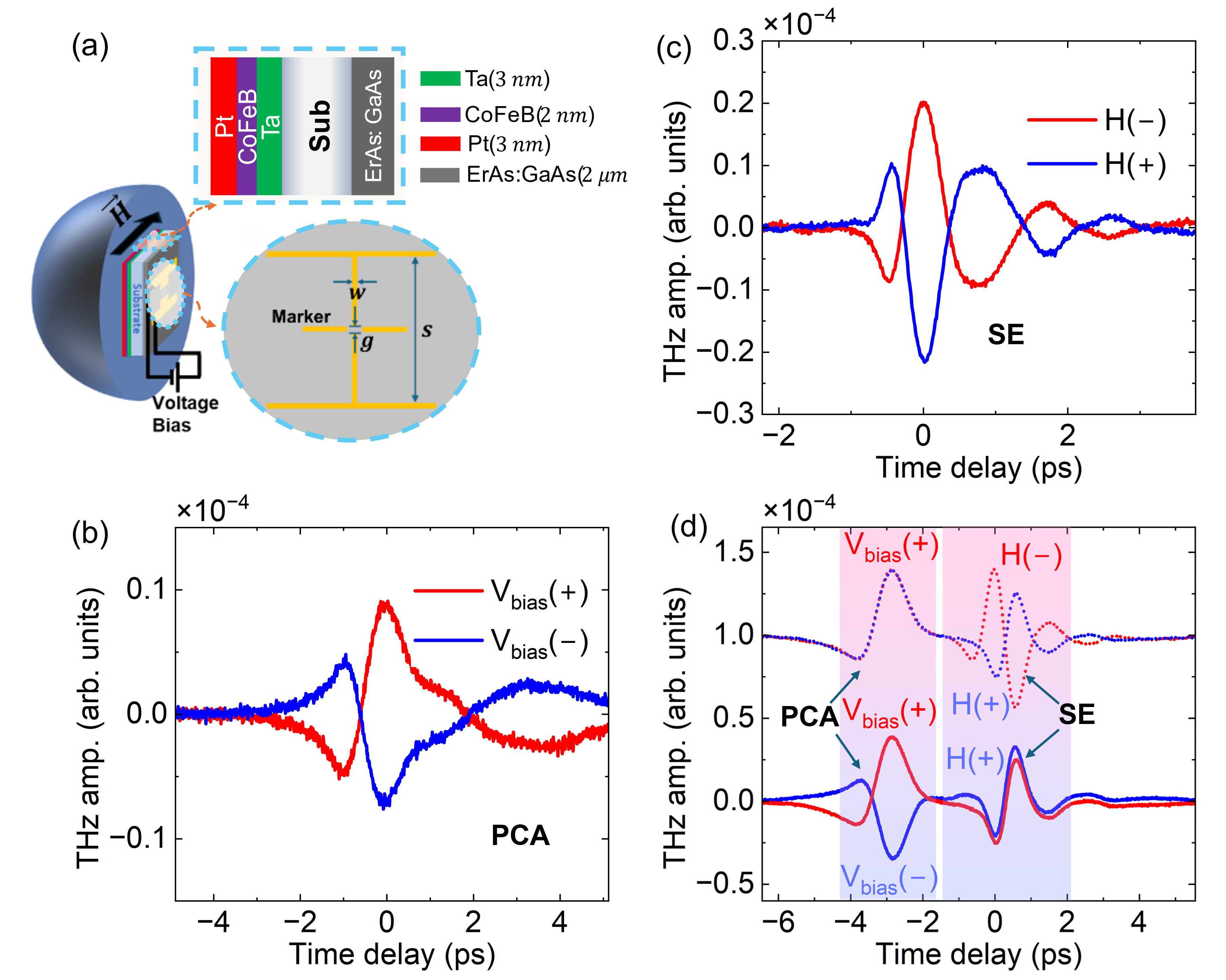}
    \caption{(a) Schematic diagram of the hybrid emitter with the antenna design, including the definition of the dimensions, the stacking order, and the thickness of each layer.  {(b,c) Demonstration of the independent tunability of the two THz sources integrated into a single on-chip: (b) time-domain THz signal when the 1560 nm is blocked and only the PCA contributes to the signal (here shown: perpendicuar configuration). (c) Corresponding time-domain THz signal when the 780 nm beam is blocked and only the SE emits a signal (here shown: perpendicular configuration.} {(d) Demonstration of the detected THz radiation from the two THz sources with simultaneous excitation at two wavelengths in the parallel configuration: the bias voltage ($\pm 4$ V) is applied in the x0-direction and the magnetic field ($\pm 1000$ Oe) is applied in the y0-direction where the emitted THz radiations from the PCA and the spintronic emitter are polarized in the $x0$-direction.} 
    }
    \label{fig2}
\end{figure}
\newpage

\begin{figure*}[t]
    \centering
    \includegraphics[width=1\columnwidth]{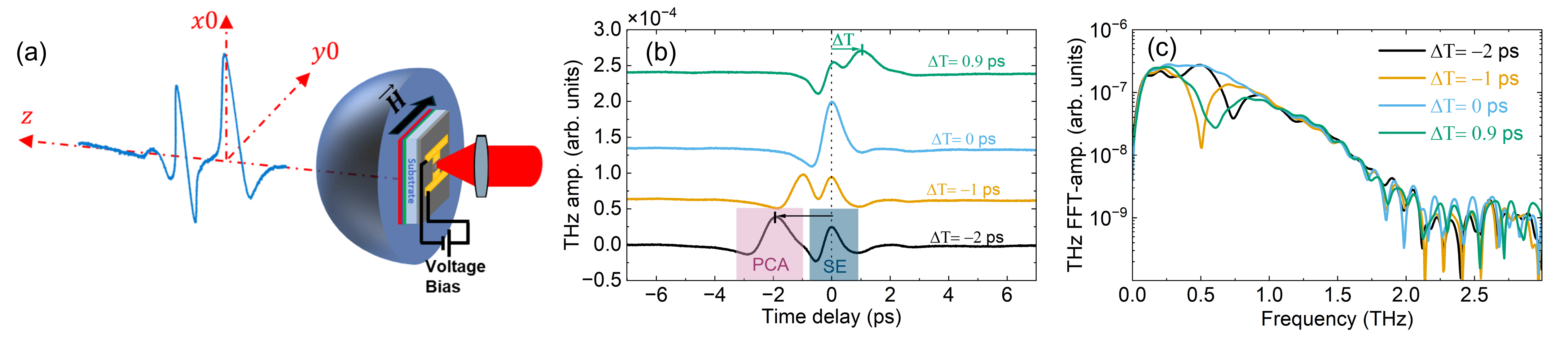}    
    \caption{(a) Schematic illustration of the hybrid emitter in the parallel configuration where the magnetic field $H$ is in the $-y0$-direction (leading to a THz electric field polarized in the $x0$-direction) and an applied bias voltage is in the $x0$-direction (also leading to a THz electric field polarized in the $x0$-direction). (b) The measured THz pulses from the PCA emitter (light red shaded area) and spintronic emitter ({SE}) (light blue shaded area) at different relative time delays ({$\Delta T$}) between the peak position. (c) The corresponding Fourier-transformed spectra of the THz pulses in the frequency domain.}
    \label{fig3}
\end{figure*}

\clearpage
\begin{figure*}[t]
    \centering
    \includegraphics[width=1\columnwidth]{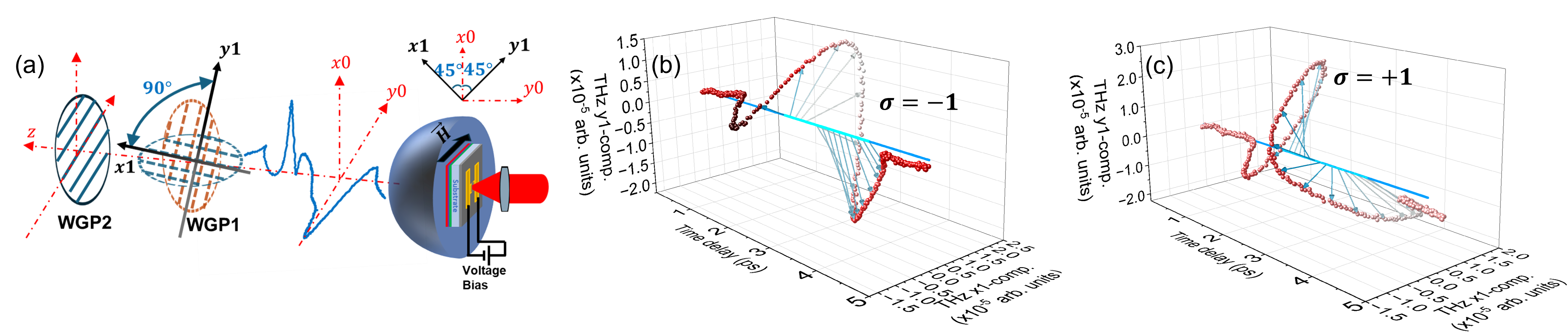}
    \caption{(a) Schematic illustration of the experimental configuration for a perpendicular alignment, which enables polarization control of the THz emission. Here, the magnetic field and the applied bias voltage are both in the $-y0$-direction. This results in an electric field polarization from the spintronic emitter in the $x0$-direction, while the corresponding emission from the PCA is polarized in the $-y0$-direction. A pair of wire-grid polarizers (WGP1 and WGP2) is used to reconstruct the polarization from the two emitters. The inset shows the relationship between $x0/y0$ and $x1/y1$ coordinate systems defined by the emission from the hybrid source and the configuration of the two wire-grid polarizers. (b) and (c) show two representative THz waveforms with different chiralities ($\sigma = \pm 1$): (b) left-handed and (c) right-handed elliptical polarization.}
    \label{fig4}
\end{figure*}

\end{document}